\documentclass[twocolumn,showpacs,preprintnumbers,amssymb,nofootinbib]{revtex4}

\usepackage{graphicx}
\usepackage{bm} 

\begin{document}


\title{Small Kerr$-$anti-de Sitter black holes are unstable}

\author{Vitor Cardoso}
\email{vcardoso@teor.fis.uc.pt} \affiliation{Centro de F\'{\i}sica
Computacional, Universidade de Coimbra, P-3004-516 Coimbra,
Portugal}
\author{\'Oscar J. C. Dias}
\email{odias@ualg.pt} \affiliation{ Centro
Multidisciplinar de Astrof\'{\i}sica - CENTRA, Departamento de
F\'{\i}sica, F.C.T., Universidade do Algarve, Campus de Gambelas,
8005-139 Faro, Portugal}

\date{\today}

\begin{abstract}
Superradiance in black hole spacetimes can trigger
instabilities. Here we show that, due to superradiance,
small Kerr-anti-de Sitter black holes are unstable.
Our demonstration uses a matching procedure, 
in a long wavelength approximation.

\end{abstract}


\maketitle
\newpage
\section{Introduction}
Einstein equations describing General Relativity and Gravitation, form a 
system of coupled non-linear partial differential equations, which is 
extremely hard to solve, even resorting to state-of-the-art computing. 
Therefore exact solutions to Einstein equations, which are possible to 
obtain only in special instances, are of fundamental importance. They allow 
us to probe essential features of General Relativity. For example, having at 
hand the spherically symmetric Schwarzschild solution it was possible to 
match General Relativity predictions against experimental observations. Once 
an exact solution is found, one must examine it in detail, and investigate 
the physical properties of such a solution. One of the most important 
aspects is the stability of a given solution. In fact, if a solution is not 
stable, then it will most certainly not be found in nature, unless the 
instability timescale is much larger than the age of our universe. 
What does one mean by stability?  
In this classical context, stability means that a 
given initially  bounded perturbation of the spacetime remains bounded for 
all times.   
For example, the Schwarzschild spacetime is stable against all kinds of
perturbations, massive or massless \cite{regge}. On the other hand the
Kerr spacetime, describing a rotating black hole, is stable against
massless field perturbations but not against massive bosonic fields
\cite{detweiler}.  

The physics behind this instability is related to a
phenomenon known as superradiance, a process which is known for
several decades, and which consists basically on a scattering process
which extracts energy from the scattering potential.  For
example, the Klein-Gordon equation for a charged scalar particle on a
step-like potential already displays such a ``superradiant''
scattering, i.e., the energy of the reflected wave is larger than the
incident one \cite{manogue,greiner}.  The first classical example of
superradiant scattering, which would lead to the notion of
superradiant scattering in black hole spacetimes, was given by
Zel'dovich \cite{zel1}, by examining what happens when scalar waves
impinge upon a rotating cylindrical absorbing object. Considering a
wave of the form $e^{-i\omega t + i m \phi}$ incident upon such a
rotating object, Zel'dovich concluded that if the frequency $\omega$
of the incident wave satisfies 
\begin{equation} \omega < m \Omega \,,
\label{super} \end{equation}
where $\Omega$ is the angular velocity of the body, then the
scattered wave is amplified.  If the ``rotating object'' is a Kerr
black hole, then superradiant scattering also occurs
\cite{zel1,bardeen,staro1} for frequencies $\omega$
satisfying (\ref{super}), but where $\Omega$ is now the angular
velocity of the black hole.  If one could find a way to
feed the amplified scattered wave onto the black hole again, then
one could in principle extract as much energy as one likes from
the black hole (as long as it is less than the total rotational
energy). The first proposal of this kind was in fact made by
Zel'dovich \cite{zel1}, who suggested to surround the rotating
cylinder by a reflecting mirror. In this case the wave would
bounce back and forth, between the mirror and the cylinder, amplifying
itself each time. A similar situation can be achieved for a Kerr
black hole: surround it by a spherical mirror and excite a given
multipole $m$ wave in it. Then the total extracted energy
should grow exponentially until finally the radiation pressure
destroys the mirror. This is exactly the same principle
behind the instability of Kerr black holes against massive bosonic
perturbations, because in this case the mass of the field works as
a wall near infinity \cite{detweiler}. 

The system black hole plus mirror is known as Press and Teukolsky's black hole
bomb \cite{press}, which has been recently investigated 
in detail in \cite{blackbomb}. It was shown in \cite{blackbomb} 
that for the system to really become unstable, the mirror must have 
a radius larger than a certain critical value. This is because the
oscillation frequencies are dictated by the mirror, and go like
$1/r_0$, with $r_0$ being the mirror radius. Thus for superradiance to
work, one must have by (\ref{super}), $1/r_0 \lesssim m\Omega$.  In principle,
black holes in anti-de Sitter (AdS) space should have similar properties as
those of the black hole bomb, since the boundary of anti-de Sitter
spacetime behaves as a wall. In fact, a similar reasoning applied to
Kerr$-$anti-de Sitter (Kerr-AdS) black holes lead one to verify the 
stability of large rotating black holes in anti-de Sitter spacetime 
(the stability of these large black holes was proven by Hawking and Reall 
\cite{hawking}), and lead also to the conjecture that small Kerr-AdS 
black holes should be unstable \cite{blackbomb}.  
The purpose of the present paper is to
prove the instability of small Kerr-AdS black holes, by
solving directly the wave equation for a scalar field, in the large
wavelength approximation, by using matched asymptotic expansions.

\section{Formulation of the problem and basic equations}
\label{formulation}
We shall consider a scalar field in the vicinity of a Kerr-AdS
black hole, with an exterior geometry described by the line
element \cite{Carter}
\begin{eqnarray}
ds^2 \!\!&=&\!\! -\frac{\Delta_r}{\rho^2}\left (
dt-\frac{a}{\Sigma}\sin^2\theta \,d\phi\right )^2
+\frac{\rho^2}{\Delta_r}\,dr^2
+\frac{\rho^2}{\Delta_{\theta}}\,d\theta^2
 \nonumber \\
& &  +\frac{\Delta_\theta}{\rho^2} \sin^2\theta\left (
a\,dt-\frac{r^2+a^2}{\Sigma} \,d\phi\right )^2 \,,
 \label{metric}
\end{eqnarray}
with
\begin{eqnarray}
& & \Delta_r=\left (r^2+a^2\right )\left (1+\frac{r^2}{\ell^2}
\right )-2Mr\,, \qquad  \Sigma=1-\frac{a^2}{\ell^2}
\nonumber \\
& &\Delta_{\theta}= 1-\frac{a^2}{\ell^2}\cos^2\theta\,, \qquad
\rho^2=r^2+a^2 \cos^2\theta  \,,
 \label{metric parameters}
\end{eqnarray}
and $\ell=\sqrt{-3/\Lambda}$ is the cosmological length associated
with the cosmological constant $\Lambda$. This metric describes
the gravitational field of the Kerr black hole, with mass $M$,
angular momentum $J=M a$, and has an event horizon at $r=r_+$ (the
largest root of $\Delta_r$). A characteristic and important
parameter of a Kerr black hole is the angular velocity of its
event horizon given by
\begin{equation}
\Omega=\frac{a}{r_+^2+a^2}\left ( 1-\frac{a^2}{\ell^2} \right )
\,.
  \label{Omega}
\end{equation}
In order to avoid singularities, the black hole rotation is
constrained to be
\begin{equation}
a<\ell \,.
  \label{upper rotation}
\end{equation}
In absence of sources, which we consider to be our case, the
evolution of the scalar field is dictated by the Klein-Gordon
equation in a Kerr-AdS spacetime, $[\nabla_{\mu}\nabla^{\mu}-\xi
R-\mu^2]\Phi=0$. Here, $R=-12/\ell^2$ is the Ricci scalar of the
Kerr-AdS spacetime, $\xi$ is a coupling constant, and $\mu$ is the
mass of the scalar field. For simplicity, and without loss of
generality, we choose the value of $\xi$ and $\mu$ in order that
the Klein-Gordon equation stays simply as
$\nabla_{\mu}\nabla^{\mu}\Phi=0$. To make the whole problem more
tractable, it is convenient to separate the field as \cite{brill}
\begin{equation}
\Phi(t,r,\theta,\phi)=e^{-i\omega t + i m \phi} \tilde{S}^m
_l(\theta) R(r)\,, \label{separation}
\end{equation}
where $\tilde{S}^m _l(\theta)$ are the AdS spheroidal angular
functions, 
and the azimuthal number $m$ takes on
integer (positive or negative) values. For our purposes, it is
enough to consider positive $\omega$'s in (\ref{separation})
\cite{bardeen}. Inserting this in Klein-Gordon equation, we get
the following angular and radial wave equations for
$\tilde{S}^m _l(\theta)$ and $R(r)$,
\begin{eqnarray}
& & \hspace{-0.5cm} \frac{\Delta_{\theta}}{\sin
\theta}\partial_{\theta}\left ( \Delta_{\theta} \sin \theta
\partial_{\theta} \tilde{S}^m _l \right )  \nonumber \\
& &\hspace{0.5cm}+ \left [  a^2 \omega^2 \cos^2 \theta-\frac{m^2
\Sigma^2}{\sin ^2{\theta}}+A_{lm} \Delta_{\theta} \right
]\tilde{S}^m _l =0\,, \label{wave eq separated1}
\\
& & \hspace{-0.5cm}  \Delta_r\partial_r \left ( \Delta_r
\partial_r R \right )+ {\bigl [} \omega^2(r^2+a^2)^2-2M a m
\omega r +a^2 m^2 \nonumber \\
& & \hspace{3.2cm} - \Delta_r (a^2\omega^2+ A_{lm}) {\bigr ]}
R=0\,,
 \label{wave eq separated}
\end{eqnarray}
where $A_{lm}$ is the separation constant that allows the split of
the wave equation, and is found as an eigenvalue of
 (\ref{wave eq separated1}). For small $a \omega$ and for small $a/\ell$,
the regime we shall be interested on in the next section, one has
\cite{seidel}
\begin{eqnarray}
A_{lm}=l(l+1)+{\cal O}(a^2\omega^2, a^2/\ell^2)\,.
 \label{eigenvalues}
\end{eqnarray}
The boundary conditions that one must impose upon the scalar field are 
the following. First, we require that the scalar field vanishes at
$r\rightarrow \infty$ because the AdS space behaves effectively as
a reflecting box, i.e., the AdS infinity works as a mirror wall
(but see also \cite{moss} and references therein for another possible 
set of boundary conditions).
Second, near the horizon $r=r_+$, the scalar field as given by
(\ref{separation}) behaves as
\begin{equation}
\Phi \sim e^{-i\omega t}e^{\pm i(\omega-m\Omega)
r_*}\,\,,\,\,r\rightarrow r_+\,, \label{bc2}
\end{equation}
where the tortoise $r_*$ coordinate is defined implicitly by
$dr_*/dr=(r^2+a^2)/\Delta_r$. 
Requiring ingoing waves at the horizon, which is the physically
acceptable solution, one must impose a negative group velocity $
v_{\rm gr}$ for the wave packet. Since $v_{\rm gr}=\pm 1$ we must
thus use the minus sign in (\ref{bc2}).
To satisfy these two boundary conditions simultaneously, the
frequencies $\omega$ must take on certain special values, which are
called quasinormal frequencies (QN frequencies, $\omega_{QN}$) and the
associated modes are called quasinormal modes (QNMs). In general,
$\omega_{QN}$ will be a complex quantity, signaling the decay of the
field, or then its growth. Note that according to the field
decomposition (\ref{separation}) if the imaginary part of $\omega$ is
positive then the field will grow exponentially as time goes by. Thus
we say that the system is unstable if the imaginary part of
$\omega_{QN}$ is positive.

\section{\label{sec:instability}Analytical calculation of the unstable modes}

In this section, we will show that small Kerr-AdS black holes are 
unstable. We shall, within some
approximations, compute the characteristic QN frequencies for a scalar field,
and show that they do have positive imaginary parts. 
The instability is due to
the presence of an effective ``reflecting mirror'' at the AdS infinity, since 
the waves are then
successively impinging on the small AdS black hole and being reflected at infinity
\cite{blackbomb}. We shall see that this interpretation agrees in all aspects
with the study of the black hole bomb \cite{blackbomb}.

We assume that $1/\omega \gg M$, i.e., that the Compton wavelength of
the scalar particle is much larger than the typical size of the black
hole, and that the AdS black hole is small, i.e., that the size of the
black hole is much smaller than the typical AdS radius, $r_+/\ell \ll
1$ . We will also assume slow rotation: $a\ll M$, and $a\ll
\ell$. Following a matching procedure introduced in
\cite{staro1,maldacstroming,unruh}, we divide the space outside the
event horizon in two regions, namely, the near-region, $r-r_+ \ll
1/\omega$, and the far-region, $r-r_+ \gg M$. We will solve the radial
equation (\ref{wave eq separated}) in each one of these two
regions. Then, we will match the near-region and the far-region
solutions in the overlapping region where $M \ll r-r_+ \ll 1/\omega$
is satisfied. When the correct boundary conditions are imposed upon
the solutions, we shall get a defining equation for $\omega_{QN}$, and 
the stability or instability of
the spacetime depends basically on the sign of the imaginary component
of $\omega_{QN}$.

\subsection{\label{sec:BH Near region}Near-region wave equation and solution}
For small AdS black holes, $r_+/\ell \ll 1$, in the near-region,
$r-r_+ \ll 1/\omega$, we can neglect the effects of the
cosmological constant, $\Lambda\sim 0$. Moreover, one has $r\sim
r_+$, $r_+ \sim 2M$, and $\omega a^2 \sim 0$ (since $\omega\ll
M^{-1}$ and $a\ll M$), and $\Delta_r \sim \Delta$ with
\begin{eqnarray}
\Delta=r^2+a^2-2Mr\,.
 \label{near wave eqDelta}
\end{eqnarray}
The near-region radial wave equation can then be written as
\begin{eqnarray}
\Delta\partial_r \left ( \Delta\partial_r R \right )+ r_+^4
(\omega-m\Omega)^2\,R - l(l+1)\Delta\,R=0 \, .
 \label{near wave eq}
\end{eqnarray}
To find the analytical solution of this equation, one first
introduces a new radial coordinate,
\begin{eqnarray}
z=\frac{r-r_+}{r-r_-}\, , \qquad 0\leq z \leq 1\,,
 \label{new radial coordinate}
\end{eqnarray}
with the event horizon being at $z=0$. Then, one has $\Delta
\partial_r=(r_+-r_-)z\partial_z$, and the near-region radial wave
equation can be written as
\begin{eqnarray}
& & \hspace{-0.5cm} z(1\!-\!z)\partial_z^2 R+ (1\!-\!z)\partial_z
R+ \varpi^2
\frac{1\!-\!z}{z} R - \frac{l(l+1)}{1\!-\!z} R=0\,, \nonumber \\
& &
 \label{near wave eq with z}
\end{eqnarray}
where we have defined the superradiant factor
\begin{eqnarray}
\varpi \equiv(\omega-m\Omega)\frac{r_+^2}{r_+-r_-} \,.
 \label{superradiance factor}
\end{eqnarray}
Through the definition
\begin{eqnarray}
R=z^{i \,\varpi} (1-z)^{l+1}\,F \,,
 \label{hypergeometric function}
\end{eqnarray}
the near-region radial wave equation becomes
\begin{eqnarray}
& &  \hspace{-0.5cm} z(1\!-\!z)\partial_z^2 F+ {\biggl [} (1+i\,
2\varpi)-\left [ 1+2(l+1)+ i\, 2\varpi \right ]\,z {\biggr ]}
\partial_z F \nonumber \\
& & \hspace{1.5cm}- \left [ (l+1)^2+ i \,2\varpi (l+1)\right ]
F=0\,.
 \label{near wave hypergeometric}
\end{eqnarray}
This wave equation is a standard hypergeometric equation
\cite{abramowitz}, $z(1\!-\!z)\partial_z^2
F+[c-(a+b+1)z]\partial_z F-ab F=0$, with
\begin{eqnarray}
& & \hspace{-0.5cm} a=l+1+i\,2\varpi \,,  \qquad b=l+1 \,, \qquad
c=1+ i\,2\varpi \,, \nonumber \\
& &
 \label{hypergeometric parameters}
\end{eqnarray}
and its most general solution in the neighborhood of $z=0$ is $A\,
z^{1-c} F(a-c+1,b-c+1,2-c,z)+B\, F(a,b,c,z)$. Using
(\ref{hypergeometric function}), one finds that the most general
solution of the near-region equation is
\begin{eqnarray}
 \hspace{-0.5cm} R &=& A\, z^{-i\,\varpi}(1-z)^{l+1}
F(a-c+1,b-c+1,2-c,z)\nonumber \\
& & +B\,z^{i\,\varpi}(1-z)^{l+1} F(a,b,c,z) \,.
 \label{hypergeometric solution}
\end{eqnarray}
The first term represents an ingoing wave at the horizon $z=0$,
while the second term represents an outgoing wave at the horizon.
We are working at the classical level, so there can be no outgoing
flux across the horizon, and thus one sets $B=0$ in
(\ref{hypergeometric solution}). One is now interested in the
large $r$, $z\rightarrow 1$, behavior of the ingoing near-region
solution. To achieve this aim one uses the $z \rightarrow 1-z$
transformation law for the hypergeometric function
\cite{abramowitz},
\begin{eqnarray}
& \hspace{-2cm} F(a\!-\!c\!+\!1,b\!-\!c\!+\!1,2\!-\!c,z)=
(1\!-\!z)^{c-a-b}  & \nonumber \\
&\times
\frac{\Gamma(2-c)\Gamma(a+b-c)}{\Gamma(a-c+1)\Gamma(b-c+1)}
 \,F(1\!-\!a,1\!-\!b,c\!-\!a\!-\!b\!+\!1,1\!-\!z) & \nonumber \\
&  \hspace{-0.2cm}+
\frac{\Gamma(2-c)\Gamma(c-a-b)}{\Gamma(1-a)\Gamma(1-b)}
 \,F(a\!-\!c\!+\!1,b\!-\!c\!+\!1,-c\!+\!a\!+\!b\!+\!1,1\!-\!z), & \nonumber \\
&
 \label{transformation law}
\end{eqnarray}
and the property $F(a,b,c,0)=1$. Finally, noting that when
$r\rightarrow \infty$ one has $1-z = (r_+-r_-)/r$, one
obtains the large $r$ behavior of the ingoing wave solution in the
near-region,
\begin{eqnarray}
R &\sim& A\,\Gamma(1-i\,2\varpi){\biggl [}
\frac{(r_+-r_-)^{-l}\,\Gamma(2l+1)}{\Gamma(l+1)\Gamma(l+1-i\,2\varpi)}\:
r^{l}\nonumber \\
& &
 +\frac{(r_+-r_-)^{l+1}\,\Gamma(-2l-1)}{\Gamma(-l)\Gamma(-l-i\,2\varpi)}\: r^{-l-1}
{\biggr ]}.
 \label{near field-large r}
\end{eqnarray}
\subsection{\label{sec:BH Far region}Far-region wave equation and solution}

In the far-region, $r-r_+ \gg M$, the effects induced by the black
hole can be neglected ($a\sim 0$, $M \sim 0$, $\Delta_r \sim
r^2[1+r^2/\ell^2]$) and the radial wave equation
 (\ref{wave eq separated}) reduces to the wave equation of a
scalar field of frequency $\omega$ and angular momentum $l$ in a
pure AdS background,
\begin{eqnarray}
& & \hspace{-1cm}(r^2+\ell^2)\partial_r^2 R+ 2\left
(2r+\frac{\ell^2}{r}\right
)\partial_r R \nonumber \\
& & +\ell^2 \left [
\omega^2\frac{\ell^2}{r^2+\ell^2}-\frac{l(l+1)}{r^2}\right ]R=0\,.
 \label{far wave eq}
\end{eqnarray}
Notice that in the above approximation, the far-region wave
equation in the Kerr-AdS black hole background is equal to the
wave equation in the pure AdS background. However, one must be cautious   
since the boundaries of the far-region in the
Kerr-AdS black hole case are $r=r_+$ and $r=\infty$, while in the
pure AdS case the boundaries are $r=0$ and $r=\infty$. In what
follows we will find the solution of (\ref{far wave eq}), first in
the pure AdS case, and then we will use this last solution to find
the far-region solution of the Kerr-AdS black hole case.

The wave equation (\ref{far wave eq}) can be written in a standard
hypergeometric form. First we introduce a new radial coordinate,
\begin{eqnarray}
x=1+\frac{r^2}{\ell^2}\, , \qquad 1\leq x \leq \infty\,,
 \label{new far radial coordinate}
\end{eqnarray}
with the origin of the AdS space, $r=0$, being at $x=1$, and
$r=\infty$ corresponds to $x=\infty$. Then, one has
$\partial_r=2\ell^{-1}\sqrt{x-1}\partial_x$, and the radial wave
equation can be written as
\begin{eqnarray}
& & \hspace{-0.5cm} x(1\!-\!x)\partial_x^2 R+ \frac{2-5 x}{2}
\,\partial_x R- \!\left [ \frac{\omega^2 \ell^2}{4x} R +
\frac{l(l+1)}{4(1\!-\!x)}\right ]\! R=0\,.
\nonumber \\
& &
 \label{far wave eq with x}
\end{eqnarray}
Through the definition
\begin{eqnarray}
R=x^{\omega \ell/2} (1-x)^{l/2}\,F \,,
 \label{far hypergeometric function}
\end{eqnarray}
the radial wave equation becomes
\begin{eqnarray}
& &  \hspace{-0.5cm} x(1\!-\!x)\partial_x^2 F+ {\biggl [}
(1+\omega\ell)-\left ( l+\frac{5}{2}+\omega\ell\right )\,x {\biggr
]}
\partial_x F \nonumber \\
& & \hspace{1.5cm}- \frac{1}{4}(l+\omega\ell)(l+3+\omega\ell)
F=0\,.
 \label{far wave hypergeometric}
\end{eqnarray}
This wave equation is a standard hypergeometric equation
\cite{abramowitz}, $x(1\!-\!x)\partial_x^2
F+[\gamma-(\alpha+\beta+1)x]\partial_x F-\alpha\beta F=0$, with
\begin{eqnarray}
& & \hspace{-0.5cm} \alpha=\frac{l+3+\omega\ell}{2} \,, \qquad
\beta=\frac{l+\omega\ell}{2} \,, \qquad
\gamma=1+\omega\ell \,, \nonumber \\
& &
 \label{far hypergeometric parameters}
\end{eqnarray}
and its most general solution in the neighborhood of $x=\infty$ is
$C\, x^{-\alpha}
F(\alpha,\alpha-\gamma+1,\alpha-\beta+1,1/x)+D\,x^{-\beta}
F(\beta,\beta-\gamma+1,\beta-\alpha+1,1/x)$. Using
 (\ref{far hypergeometric function}), one finds that the most general
solution for $R(x)$ is
\begin{eqnarray}
 R \!\!&=&\!\! C\, x^{-(l+3)/2}(1-x)^{l/2}
F(\alpha,\alpha-\gamma+1,\alpha-\beta+1,1/x) \nonumber \\
& & +D\,x^{-l/2}(1-x)^{l/2}
F(\beta,\beta-\gamma+1,\beta-\alpha+1,1/x) \,. \nonumber \\
& &
 \label{far hypergeometric solution}
\end{eqnarray}
Since $F(a,b,c,0)=1$, as $x\rightarrow \infty$ this solution
behaves as $R\sim (-1)^{l/2}(C x^{-2}+D)$. But the AdS infinity
behaves effectively as a wall, and thus the scalar field must
vanish there which implies that we must set $D=0$ in
 (\ref{far hypergeometric solution}).
 We are now interested in the small $r$,
$x\rightarrow 1$, behavior of (\ref{far hypergeometric solution}).
To achieve this aim one uses the $1/x \rightarrow 1-x$
transformation law for the hypergeometric function
\cite{abramowitz},
\begin{eqnarray}
&  \hspace{-0.5cm}
F(\alpha,\alpha\!-\!\gamma\!+\!1,\alpha\!-\!\beta\!+\!1,1/x)=
x^{\alpha-\gamma+1}(x-1)^{\gamma-\alpha-\beta}  & \nonumber \\
&\times
\frac{\Gamma(\alpha-\beta+1)\Gamma(\alpha+\beta-\gamma)}{\Gamma(\alpha)\Gamma(\alpha-\gamma+1)}
 \,F(1\!-\!\beta,1\!-\!\alpha,\gamma\!-\!\alpha\!-\!\beta\!+\!1,1\!-\!x)& \nonumber \\
&+ \,x^\alpha
\frac{\Gamma(\alpha-\beta+1)\Gamma(\gamma-\alpha-\beta)}{\Gamma(1-\beta)\Gamma(\gamma-\beta)}
 \,F(\alpha,\beta,\alpha\!+\!\beta\!-\!\gamma\!+\!1,1\!-\!x),  & \nonumber \\
 & &
 \label{far transformation law}
\end{eqnarray}
and the property $F(a,b,c,0)=1$. Finally, noting that when
$x\rightarrow 1$ one has $x-1\rightarrow r^2/\ell^2$, one obtains
the small $r$ behavior of $R(r)$,
\begin{eqnarray}
& & \hspace{-1.5cm} R \sim C\,\Gamma(5/2){\biggl [}
\frac{(-1)^{l/2}\ell^{-l}\,\Gamma(-l-\frac{1}{2})}{\Gamma(1-\frac{l}{2}-\frac{\omega\ell}{2})
\Gamma(1-\frac{l}{2}+\frac{\omega\ell}{2})}\:
r^{l}\nonumber \\
& &
 +\frac{(-1)^{-3l/2}\ell^{l+1}\,\Gamma(l+\frac{1}{2})}
 {\Gamma(\frac{3}{2}+\frac{l}{2}+\frac{\omega\ell}{2})
\Gamma(\frac{3}{2}+\frac{l}{2}-\frac{\omega\ell}{2})}\: r^{-l-1}
{\biggr ]}.
 \label{far field-small r}
\end{eqnarray}

The boundaries of the pure AdS spacetime are the origin, $r=0$,
and the effective wall at $r=\infty$. When $r\rightarrow 0$, the
wave solution $R$ diverges, since $r^{-l-1}\rightarrow \infty$ in
(\ref{far field-small r}). In order to have a regular solution at
the origin we must then demand that
$\Gamma(\frac{3}{2}+\frac{l}{2}-\frac{\omega\ell}{2})=\infty$.
This occurs when the argument of the gamma function is a
non-positive integer, $\Gamma(-n)=\infty$ with $n=0,1,2,\cdots$.
Therefore, the requirement of regularity of the wave solution at
the origin selects the frequencies that might propagate in the AdS
background. These are given by the discrete spectrum $\omega
\ell=l+3+2n$, which agrees with known results \cite{cardoso,burgess}. 
We remark that, alternatively, in order to have a
regular solution at the origin we could have required
$\Gamma(\frac{3}{2}+\frac{l}{2}+\frac{\omega\ell}{2})=\infty$.
This option would lead to the negative spectrum $\omega
\ell=-(l+3+2n)$, which of course must also be a solution.
However, to simplify matters we shall only deal with positive frequencies,
as was said earlier.

Now that we have found the wave solution that propagates in a pure
AdS spacetime, we can discuss the far-region solution in a
Kerr-AdS background. As we pointed out earlier, the main difference
between the two solutions lies on the inner boundary: $r=0$ in
the pure AdS case and $r=r_+$ in the black hole case. We expect
that the allowed spectrum of discrete real frequencies that can
propagate in the far-region of the Kerr-AdS black hole is equal to
the one of the pure AdS background since, at large distances from
the inner boundary, both backgrounds are similar. However, the
existence of the black hole inner boundary implies that once
radiation crosses this zone it will be scattered by
the black hole (more precisely it will be scattered by the potential 
barrier outside the event horizon) and its amplitude will decrease or, 
eventually, since conditions for superradiance might be present, it will grow
leading to an instability. Therefore, in the spirit of
\cite{detweiler}, we expect that the presence of this scattering
by the black hole induces a small complex imaginary part in the
allowed frequencies, $\delta={\rm Im}[\omega]$, that describes the
slow decay of the amplitude of the wave if $\delta<0$, or the
slowly growing instability of the mode if $\delta>0$. Summarizing,
the frequencies  that can propagate in the Kerr-AdS background are
given by
\begin{eqnarray}
\omega_{QN}=\frac{l+3+2n}{\ell}+i\delta\,,
 \label{frequency spectrum}
\end{eqnarray}
with $n$ being a non-negative integer, and $\delta$ being a small
quantity. The small $r$ behavior of the radial wave solution in
the Kerr-AdS background is described by
 (\ref{far field-small r}), subjected to the regularity condition
 (\ref{frequency spectrum}). Now, we want to extract $\delta$ from
 the gamma function in (\ref{far field-small r}). This is done in Appendix
 \ref{sec:A1}, yielding for small $\delta$ and for small $r$ the result
\begin{eqnarray}
& & \hspace{-1cm} R \sim C\,\Gamma(5/2){\biggl [}
\frac{(-1)^{l/2}\ell^{-l}\,\Gamma(-l-\frac{1}{2})}{\Gamma(
-l-\frac{1}{2}-n) \Gamma(\frac{5}{2}+n)}\:
r^{l}\nonumber \\
& & \hspace{-0.3cm}
 +i\,\delta\,\frac{\Gamma(l+1/2)}{2}\,
\frac{(-1)^{-3l/2+n+1}\ell^{l+2} n!}{(l+2+n)!} \: r^{-l-1} {\biggr
]}.
 \label{far field-small r KerrAdS}
\end{eqnarray}

\subsection{\label{sec:BH Matching}Matching conditions. Properties of the unstable modes}

When $M \ll r-r_+ \ll 1/\omega$, the near-region solution and the
far-region solution overlap, and thus one can match the large $r$
near-region solution (\ref{near field-large r}) with the small $r$
far-region solution (\ref{far field-small r KerrAdS}). This
matching yields
\begin{eqnarray}
& & \hspace{-0.4cm} \delta\simeq -2i \,
 \frac{(-1)^{n+1}\ell^{-2(l+1)}}{\Gamma(l+1/2)}
 \frac{\,(r_+-r_-)^{2l+1}}{\Gamma(n+5/2)}
 \frac{\Gamma(l+1-i\,2\varpi)} {\Gamma(-l-i\,2\varpi)}
   \nonumber \\
& &
  \times
\frac{\Gamma(-2l-1)}{\Gamma(-l)}
  \frac{\Gamma(-l-1/2)}{\Gamma(-l-1/2-n)}
 \frac{\Gamma(l+1)}{\Gamma(2l+1)} \frac{(l+2+n)!}{n!}\,.\nonumber \\
& &
 \label{aux delta}
\end{eqnarray}
Using the property of the gamma function,
$\Gamma(1+x)=x\Gamma(x)$, we can find the values of all the gamma
functions that appear in (\ref{aux delta}) yielding simply (see
Appendix \ref{sec:A2}),
\begin{eqnarray}
& & \hspace{-0.4cm} \delta\simeq -\sigma \left (
\frac{l+3+2n}{\ell}-m\Omega\right )\,
 \frac{r_+^{\,2} (r_+-r_-)^{2l}} {\pi \,\ell^{2(l+1)}}
 \,,
 \label{delta}
\end{eqnarray}
with
\begin{eqnarray}
& & \sigma\equiv \frac{(l!)^2(l+2+n)!}{(2l+1)! \, (2l)! \, n!}
  \frac{2^{l+4} \,(2l+1+n)!!}{(2l-1)!! \, (2l+1)!! \, (2n+3)!!}
  \nonumber \\
& & \hspace{1cm} \times \left (  \prod_{k=1}^l (k^2+4\varpi^2)
\right )
 \,, \nonumber \\
& &
 \label{delta parameter}
\end{eqnarray}
and $\varpi =\left ( \frac{l+3+2n}{\ell}-m\Omega \right )
\frac{r_+^2}{r_+-r_-}$. Equations (\ref{frequency spectrum}) and
(\ref{delta}) are the main results of this paper. We have,
\begin{eqnarray}
\delta\propto -\left ( {\rm Re}[\omega_{QN}]-m\Omega \right ) \,.
 \label{conclusion 1}
\end{eqnarray}
Thus, $\delta>0$ for ${\rm Re}[\omega_{QN}]<m \Omega$, and $\delta<0$
for ${\rm Re}[\omega_{QN}]>m \Omega$. The scalar field $\Phi$ has the
time dependence $e^{-i\omega t}=e^{-i{\rm Re}(\omega) t}e^{\delta
t}$ which implies that for ${\rm Re}[\omega_{QN}]<m \Omega$, the
amplitude of the field grows exponentially and the mode becomes
unstable, with a growth timescale given by $\tau=1/\delta$. This
was the main aim of this paper, namely to show that small, $r_+\ll
\ell$, Kerr-AdS black holes are unstable.
As a check of our results we note that for $l=0$ we have
$\delta\propto r_+^{\,2}$, which is in agreement with numerical
results for quasinormal modes of small Schwarzschild-anti-de Sitter
black holes \cite{horo, cardoso}.  Also, it was shown numerically in
\cite{cardoso} that for higher $l$-poles the imaginary component
decays faster with $r_+$, which is consistent with our result. Indeed,
we see from (\ref{delta}) that the imaginary part should behave as
$r_+^{2l+2}$, for non-rotating black holes.

\begin{figure} [t]
\includegraphics*[height=2.2in]{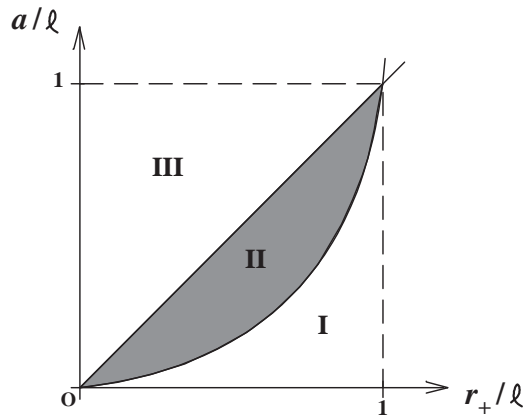}
\caption{Range of black hole parameters for which one has stable and unstable modes.
Regularity condition implies that $a/\ell<1$, and for small
Kerr-AdS black holes we have $r_+/\ell<1$. Region I represents a
stable mode zone, while regions II and III represent black holes
that can have unstable modes. To be accurate, in the
approximations we used, we can only guarantee the presence of an
instability in region II. There is however no reason to doubt that
the instability also exists in Region III. The frontier between
regions I and II is the parabola $a/\ell=r_+^{\,2}/\ell^2$. 
To ascertain the complete instability zone, numerical work is needed. }
\label{fig:instability}
\end{figure}

At this point it is appropriate to discuss the domain of validity
of our results. Our final result (\ref{frequency spectrum}) says
that ${\rm Re} [\omega_{QN}]\sim 1/\ell$, and the condition for
superradiance is ${\rm Re} [\omega_{QN}]\lesssim \Omega$. Now, we have
$\Omega \sim a/r_+^{\,2}$ in the slow rotation approximation.
Therefore, the superradiance condition together with
(\ref{frequency spectrum}) implies that the rotation parameter
must satisfy $\frac{a}{\ell} \gtrsim \frac{r_+^{\,2}}{\ell^2}$,
where the small black hole condition implies $r_+/\ell\ll 1$. This
sets the lower bound on $a/\ell$ for which instability sets in. 
The upper bound is fixed by the slow rotation
approximation $a\ll r_+$ that we used to derive our results.
Thus, within all our approximations, we see that instability sets in
for  $\frac{r_+^{\,2}}{\ell^2} \lesssim \frac{a}{\ell}
\ll \frac{r_+}{\ell}$. There is however no reason to doubt that
the instability exists all the way up
to the maximal rotation case $a=\ell$. This discussion is summarized in Fig.
\ref{fig:instability}.

\section{Conclusions}

We have shown that small, $r_+\ll \ell$, Kerr-AdS black holes are
unstable against the scattering of a wave that satisfies the
superradiant regime, $\omega<m\Omega$. This possibility was raised
in \cite{hawking}, and heuristic arguments that favored this
hypothesis were presented in \cite{blackbomb}. We have achieved
this result by analytical means in the long wavelength limit,
$\omega\ll 1/r_+$, and in the slow rotation regime, $a\ll \ell$
and $a\ll r_+$. We have provided analytical estimates for growing
timescales and oscillation frequencies of the corresponding
unstable modes. Although we have worked only with zero spin
(scalar) waves, we expect that the general features for other
spins will be the same. As shown in \cite{hawking}, large Kerr-AdS
black holes are stable.

The properties of the instabilities present in the small Kerr-AdS
black hole and in the black hole-mirror system (proposed in
\cite{press} and studied in detail in \cite{blackbomb}) are quite
similar. The black hole-mirror system, also known as
Press-Teukolsky's black hole bomb \cite{press}, consists of a Kerr
black hole in an asymptotically flat background surrounded by a
mirror placed at constant $r$, $r=r_0$. Superradiant scattering
occurs naturally in the Kerr black hole and, in this regime, if
one surrounds the black hole by a reflecting mirror, the wave will
bounce back and forth between the mirror and the black hole,
amplifying itself each time and leading to an instability. The
analogy between this system and the Kerr-AdS black hole is clear.
The AdS space behaves effectively as a box, i.e., the AdS 
wall with typical radius $\ell$ plays in this analogy the role of a mirror
wall with radius $r_0\equiv\ell$. Indeed, in the Press-Teukolsky's
black hole bomb the real part of the allowed frequency is
proportional to the inverse of the mirror's radius
\cite{blackbomb}, ${\rm Re}[\omega]\propto 1/r_0$, while in the
small Kerr-AdS black hole case we have found that ${\rm
Re}[\omega]\propto 1/\ell$. Moreover, in the Press-Teukolsky's
system the growth timescale of the instability satisfies
\cite{blackbomb} $\delta^{-1}=1/{\rm Im}[\omega]\propto
r_0^{\,2(l+1)}$, while in the AdS black hole we have
$\delta^{-1}\propto \ell^{2(l+1)}$.
Although we worked in the four dimensional case only, the general
arguments that pointed to the existence of this instability allow
one to predict that higher dimensional small Kerr-AdS black holes
are also unstable.

\section*{Acknowledgements}
The authors acknowledge Jos\'e P.S. Lemos for a critical reading of
the manuscript. This work was partially funded by Funda\c c\~ao 
para a Ci\^encia e Tecnologia (FCT) -- Portugal through 
project CERN/FNU/43797/2001. The authors acknowledge financial support 
from FCT through grant SFRH/BPD/2003.


\vskip 3mm
\appendix
\section{\label{sec:A1}The small ${\bm r}$ behavior of the far-region
solution}

In this Appendix we present the main steps that allow us to go
from (\ref{far field-small r}) into
 (\ref{far field-small r KerrAdS}). In order to do
 so, one first notes that use of (\ref{frequency spectrum}) yields
\begin{eqnarray}
& & \hspace{-1cm} \Gamma\left (
\frac{3}{2}+\frac{l}{2}+\frac{\omega\ell}{2}\right )
\Gamma\left (\frac{3}{2}+\frac{l}{2}-\frac{\omega\ell}{2}\right )\nonumber \\
 & & \hspace{0.3cm} =
\Gamma(l+3+2n+i\ell\delta/2 )\Gamma(-n-i \ell\delta/2)\,.
\end{eqnarray}
Using the gamma function properties \cite{abramowitz},
$\Gamma(k+z)=(k-1+z)(k-2+z)\cdots(1+z)\Gamma(1+z)$ with $k=l+3$
and $z=n+i\ell \delta/2$, and $\Gamma(z)\Gamma(1-z)=\pi /\sin(\pi
z)$ with $z=1+n+i\ell\delta/2$, one has (for $\delta\ll 1$) the
result
\begin{eqnarray}
& & \hspace{-1.5cm} \left [ \Gamma\left (
\frac{3}{2}+\frac{l}{2}+\frac{\omega\ell}{2}\right )
 \Gamma\left (\frac{3}{2}+\frac{l}{2}-\frac{\omega\ell}{2}\right )
 \right ]^{-1} \nonumber \\
 & & \hspace{1cm} \simeq i \,(-1)^{n+1} \frac{n!}{(l+2+n)!}\,\frac{\ell}{2}\, \delta\,.
 \label{appendix1}
\end{eqnarray}
Moreover, use of (\ref{frequency spectrum}) with $\delta\sim 0$
yields
\begin{eqnarray}
& & \hspace{-1cm} \Gamma\left (
1-\frac{l}{2}-\frac{\omega\ell}{2}\right )
\Gamma\left (1-\frac{l}{2}+\frac{\omega\ell}{2}\right ) \nonumber \\
 & & \hspace{1cm} \simeq \Gamma\left (
-l-\frac{1}{2}-n\right ) \Gamma\left (\frac{5}{2}+n\right )\,.
\label{appendix2}
\end{eqnarray}
Finally, inserting (\ref{appendix1}) and
 (\ref{appendix2}) into (\ref{far field-small r}) yields
 (\ref{far field-small r KerrAdS}).

\section{\label{sec:A2}Useful gamma function relations}

The transition from (\ref{aux delta}) into (\ref{delta}) is done
using only the gamma function property, $\Gamma(1+x)=x\Gamma(x)$.
Indeed, with it we can show that
\begin{eqnarray}
& &
\frac{\Gamma(l+1-i\,2\varpi)}{\Gamma(-l-i\,2\varpi)}=i\,(-1)^{l+1}2\varpi\prod_{k=1}^l
(k^2+4\varpi^2)\,, \nonumber \\
& & \frac{\Gamma(-2l-1)}{\Gamma(-l)}=(-1)^{l+1} \frac{l!}{(2l+1)!}\,, \nonumber \\
 & & \frac{\Gamma(-l-1/2)}{\Gamma(-l-1/2-n)}=(-1)^n 2^{-n}
\frac{(2l+1+n)!!}{(2l+1)!!}\,, \nonumber \\
 & & \Gamma(l+1/2)=2^{-l}(2l-1)!!\sqrt{\pi}\,, \nonumber \\
 & & \Gamma(n+5/2)=2^{-n-2}(2n+3)!!\sqrt{\pi}\,.
\label{gamma values}
\end{eqnarray}


\end{document}